\documentclass[aps, prd, showpacs, superscriptaddress, ctexart, nofootinbib]{revtex4}

\usepackage{graphicx}
\usepackage{latexsym}
\usepackage{subfigure}
\usepackage{amsmath,array}

\usepackage{color}

\def\be{\begin{equation}}
\def\ee{\end{equation}}
\def\bea{\begin{eqnarray}}
\def\eea{\end{eqnarray}}

\begin{document}

\title{Cosmic Reionization Study : Principle Component Analysis After Planck}
\author{Yang Liu}
\email{liuy92@ihep.ac.cn}
\affiliation{Theoretical Physics Division, Institute of High Energy Physics, Chinese Academy of Science, P.O.Box 918-4, Beijing 100049, P.R.China}

\author{Hong Li}
\email{hongli@ihep.ac.cn}
\affiliation{Key Laboratory of Particle Astrophysics, Institute of High Energy Physics, Chinese Academy of Science, P.O.Box 918-3, Beijing 100049, P.R.China}

\author{Si-Yu Li}
\affiliation{Theoretical Physics Division, Institute of High Energy Physics, Chinese Academy of Science, P.O.Box 918-4, Beijing 100049, P.R.China}

\author{Yong-Ping Li}
\affiliation{Theoretical Physics Division, Institute of High Energy Physics, Chinese Academy of Science, P.O.Box 918-4, Beijing 100049, P.R.China}

\author{Xinmin Zhang}
\affiliation{Theoretical Physics Division, Institute of High Energy Physics, Chinese Academy of Science, P.O.Box 918-4, Beijing 100049, P.R.China}


\begin{abstract}
The study of reionization history plays an important role in understanding the evolution of our universe. It is commonly believed that the intergalactic medium(IGM) in our universe are fully ionized today, however the reionizing process remains to be mysterious. A simple instantaneous reionization process is usually adopted in modern cosmology without direct observational evidence. However, the history of ionization fraction, $x_e(z)$ will influence CMB observables and constraints on optical depth $\tau$. With the mocked future data sets based on featured reionization model, we find the bias on $\tau$ introduced by instantaneous model can not be neglected. In this paper, we study the cosmic reionization history in a model independent way, the so called principle component analysis(PCA) method, and reconstruct $x_e (z)$ at different redshift $z$ with the data sets of Planck,  WMAP 9 years temperature and polarization power spectra, combining with the baryon acoustic oscillation(BAO) from galaxy survey and type Ia supernovae(SN) Union 2.1 sample respectively. The results show that reconstructed $x_e(z)$ is consistent with instantaneous behavior, however, there exists slight deviation from this behavior at some epoch. With PCA method, after abandoning the noisy modes, we get stronger constraints, and the hints for featured $x_e(z)$ evolution could become a little more obvious.

\end{abstract}


\maketitle

\section{Introduction}\label{Int}
The study on reionization history of our universe remains to be an open question in modern cosmology. It is commonly believed that our universe should be neutral after recombination epoch, the so called dark age. After that, the first generation of galaxies provide ultraviolet radiation so that the IGM start to be ionized from neutral phase\cite{{A. Loeb 2001},{N. Y. Gnedin 2006}}. Such phase transition of the IGM is the so called reionization history and it relates to many fundamental questions of astrophysics, since detailed reionization process depends on the formation and evolution of high energy astrophysical objects, such as mini-quasars, x-ray binaries, metal-free stars, etc., which provide the sources for reionization transition.

A lot of astronomical observations provide the information of reionization, for example, the level of ionized state at different epochs, even though we could not observe the reionizing process directly. Lymann-$\alpha$ forests, a series of absorption lines in the spectra of distant sources(quasars or galaxies) from the Lymann-$\alpha$ electron transition of the neutral hydrogen atom(HI), could be an important tracer of IGM ionizing. If there exists homogenious distribution of HI gas in the line of sight from source to observer, the Lymann-$\alpha$ forest will turn into a Gunn-Peterson (GP) trough\cite{Gunn}, so by observing GP trough, we can  map the neutral hydrogen in the IGM. From the detection of Lymann-$\alpha$ absorption lines or GP trough, we know that the universe is highly ionized at least until $z\sim6$\cite{{R.H. Becker 2001},{X. Fan 2006}}. On the other hand, the astronomical observations provide some evidence that there exists HI in the IGM at $z=7.1$\cite{{D.J. Mortlock},{J.S. Bolton}}. Those observations show that reionization should last for a period of time.

Some other experiments can put constraints on reionization history as well, for example 21cm experiments which can measure the distribution of HI($x_{\rm{HI}}$), and traces the evolution of reionziation.  There are a number of 21cm experiments, such as GMRT\cite{GMRT}, LOFAR\cite{LOFAR}, MWA\cite{MWA}, 21CMA\cite{21CMA}, and PAPER\cite{PAPER}. However, due to their low signal-to-noise ratios, none of them can give convincing results about history of reionization at current stage. Measuring the temperature of IGM can also give some constraints on the epoch of reionization\cite{S. Zaroubi 2012}. 

The cosmic microwave background radiation (CMB) provides useful information on reionization history. Once the IGM are ionized, there will be lots of electrons and the interactions between CMB photons and electrons through Thomson scattering will deform the black body distribution of CMB\cite{M. Zaldarriaga} which can be imprinted in CMB maps. $\tau$ is an important cosmological parameter in CMB for describing the post Thomson scattering effects, and $\tau$ is an integration of the reionized fraction (which labeled as $x_e(z)$), $\tau=\int_{\eta}^{\eta_0} d\eta a(\eta)n_e\sigma_T$, where $\eta \equiv \int {dt \slash a}$ is the conformal time, $\eta_0$ is the present time, $\sigma_T$ is the Thomson cross section and $n_e \propto (1+z)^3x_e(z)$ is the number density of free electrons produced by reionization. In CMB power spectra of TT and EE, on small angular scales, there is a reducing factor of $exp(-2\tau )$. For polarization power spectrum $C_l^{EE}$, it is enhanced on large angular scale since the scattering would generate extra polarization, and the enhancement has already detected from many CMB experiments\cite{{WMAP9},{Planck2015a}}. These effects describing here is only about the globally averaged reionization, the perturbation of the reionization(inhomogeneous effect) can change $C_l^{EE}$ on small angular scale, but it is very fainter than weak lensing effect\cite{O. Dore}, and we won't consider it in this project. There has being lots of CMB experiments done in recent decades\cite{{WMAP9},{Planck2015a},{ACT},{SPT}}, and the data are becoming more and more accurate. Also tight constraints are performed on the reionization optical depth, such as WMAP 9 years data\cite{WMAP9} gives $\tau = 0.089 \pm 0.014$ and $z_{re} = 10.6 \pm 1.1$, $\tau = 0.066 \pm 0.016$ and $z_{re} = 8.8_{-1.4}^{+1.7}$ from Planck temperature and lensing data\cite{Planck2015b}, where $z_{re}$ gives the epoch of the ionized fraction equals to one half.

\begin{figure}

\subfigure{\includegraphics[width=3in]{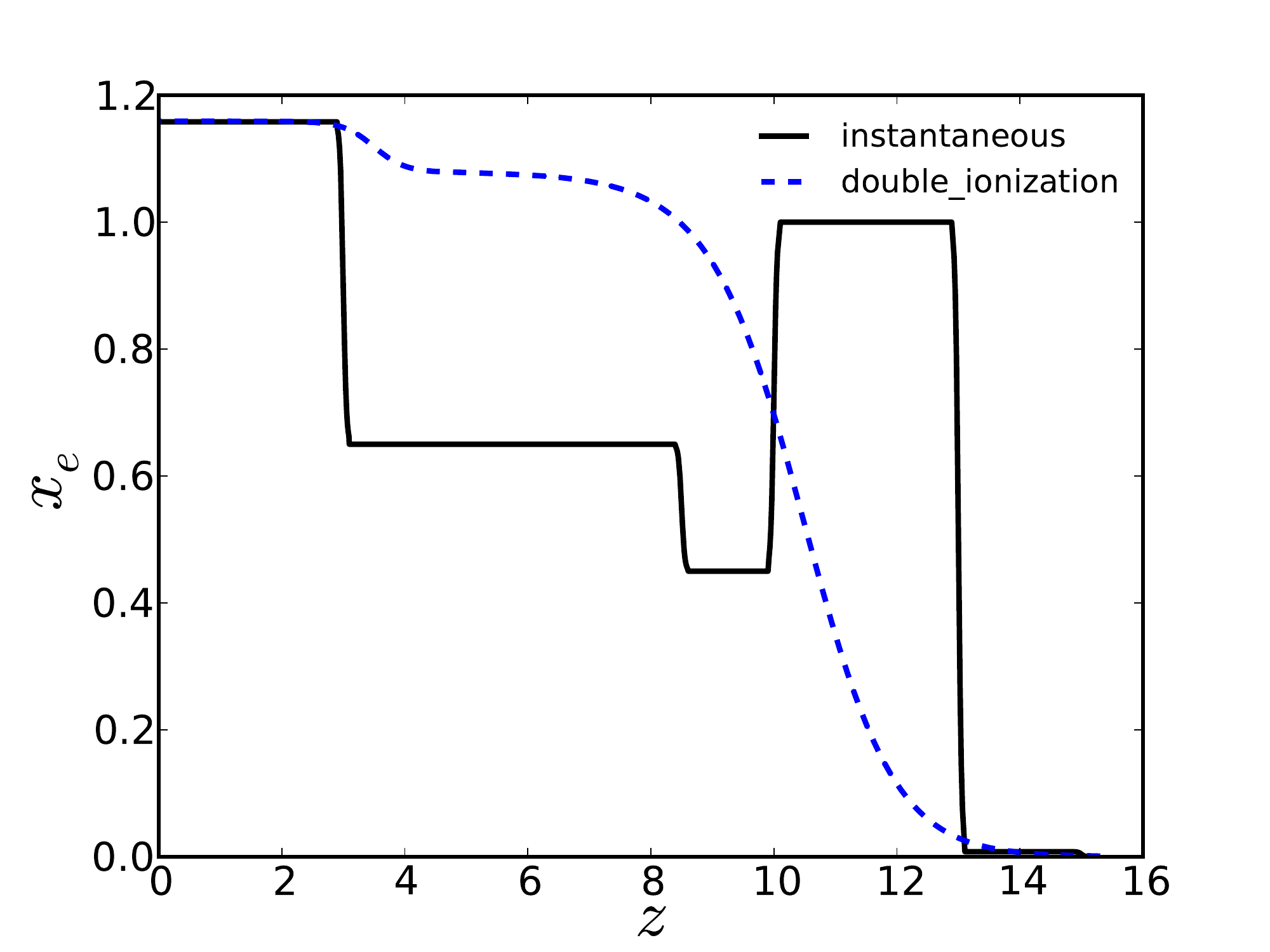}}
\subfigure{\includegraphics[width=3in]{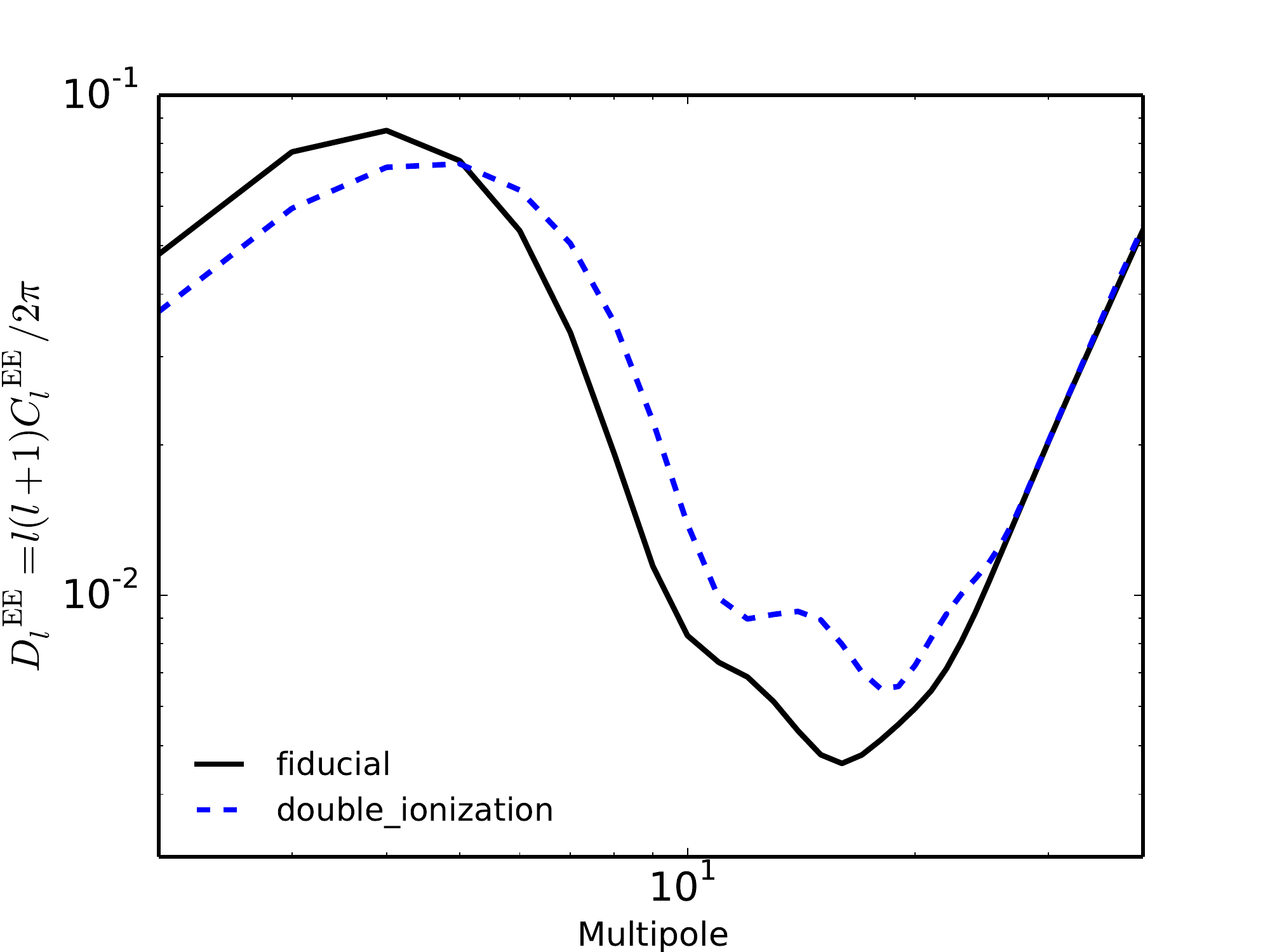}}
\caption{CMB EE power spectra(right panel) and reionization history(left panel) for two different models: instantaneous(dashed), double reionization(solid). These two models give the same optical depth($\tau = 0.085$).}\label{figure:models}

\end{figure}

When using the CMB data, people always adopt an instantaneous model to characterize the evolution of reionzation process, in which, the IGM are suddenly reionized in a very short time, and the reionizing process is very short so that the function of ionized fraction $x_e(z)$ can be described by a tanh-based function\cite{CAMB note}
\begin{equation}\label{xe_equation}
x_e(y) = \frac {f}{2} [1+ \tanh (\frac {y(z_{re})-y} {\triangle_y})],
\end{equation}
where $y\equiv (1+z)^{3\slash 2}$, and $y(z_{re})= (1+z_{re})^{3\slash 2}$ for $x_e = f\slash 2$. $f$ is a constant with value $\sim 1.08$ and $\triangle_y = 1.5\sqrt{1+z_{re}}\triangle_z$, where $\triangle_z$ is some constant, and there is one to one correspondence between the optical depth($\tau$) and the redshift of reionization($z_{re}$) when $\triangle_z$ is given. It is well known that the epoch of reionization is very complicated, there is no evidence that the history of reionization is just an instantaneous model. Also, considering that $\tau$ is the integration of the reionized fraction, the constraints on $\tau$ should be very model dependent, that is to say it will introduce bias with strong assumption on reionization model. In fact, in this way, $\tau$ can not provide more detailed information on reionization history, since different reionization models can give same optical depth. As shown in Fig. \ref{figure:models}, different reionized model could generate different $C_l^{EE}$ (in right panel) even for the same $\tau$(in the left panel). So, we see that the ionized fraction parameter $x_e(z)$ are the more basic parameters for describing reionization history.

Very recently, Planck have released their new results, the whole sky map of CMB anisotropies including both temperature and polarization, which gives the most accurate measurements on CMB. The high quality data provides a wealth of new information on cosmology. The TT spectrum is accurately measured to multipoles $l\sim 2500$, and by cross checking with the high resolution ground based CMB observations, such as ACT\cite{ACT}, SPT\cite{SPT},  the "damping tail" is measured with high accuracy in the Planck TT power spectrum at $l$ $\geq$ $2000$. The most important is that, the cross checking show that TE and EE spectra are in good agreement with TT, which provides an important test of the accuracy of the data sets.
With the new measurement, in this paper, we do a model independent study of the reionization evolution history. We totally abandon the assumption of instantaneous model, instead, we separate the reionization history into several bins in redshift space, and take $x_e(z)$ for each $z$ bin as a free parameter. By doing the global fitting, we can get constraints on $x_e(z)$ and reconstruct the reionization history. Further more, we adopt the principle component analysis (PCA) method in our study to get tighter and more reasonable constraints. PCA are widely used in the literature for optimizing the signal to noise ratio and solving for high quality estimation on target parameters.  We have noted that extracting the information of reionization history from astronomic observations are widely studied in the literatures\cite{{W. Hu 2003},{W. Hu 2008},{I. McGreer},{S. Mitra},{R.G. Cai}}. 

The structure of our paper is organized as follows. In section II
we illustrate the motivation for the model independent study of reionization history. In section III we introduce the data sets that adopted in the fitting analysis
, the global fitting procedure and principle component analysis method. The final constraints are presented in section IV. A summary and conclusion are given in section V.

\section{motivation}

\begin{figure}
\centering
   \subfigure{\includegraphics[width=3.5in]{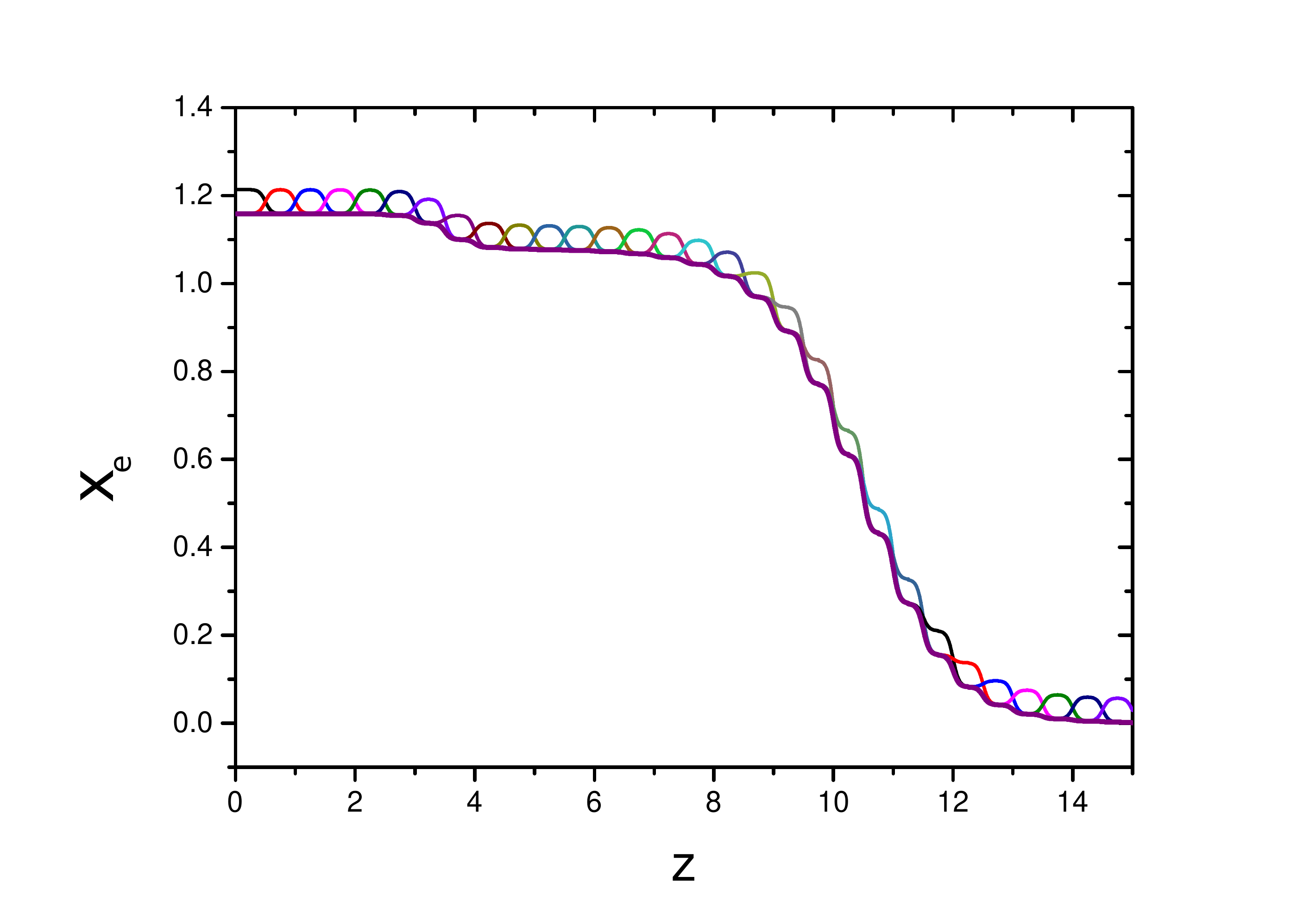}}
   \subfigure{\includegraphics[width=3.5in]{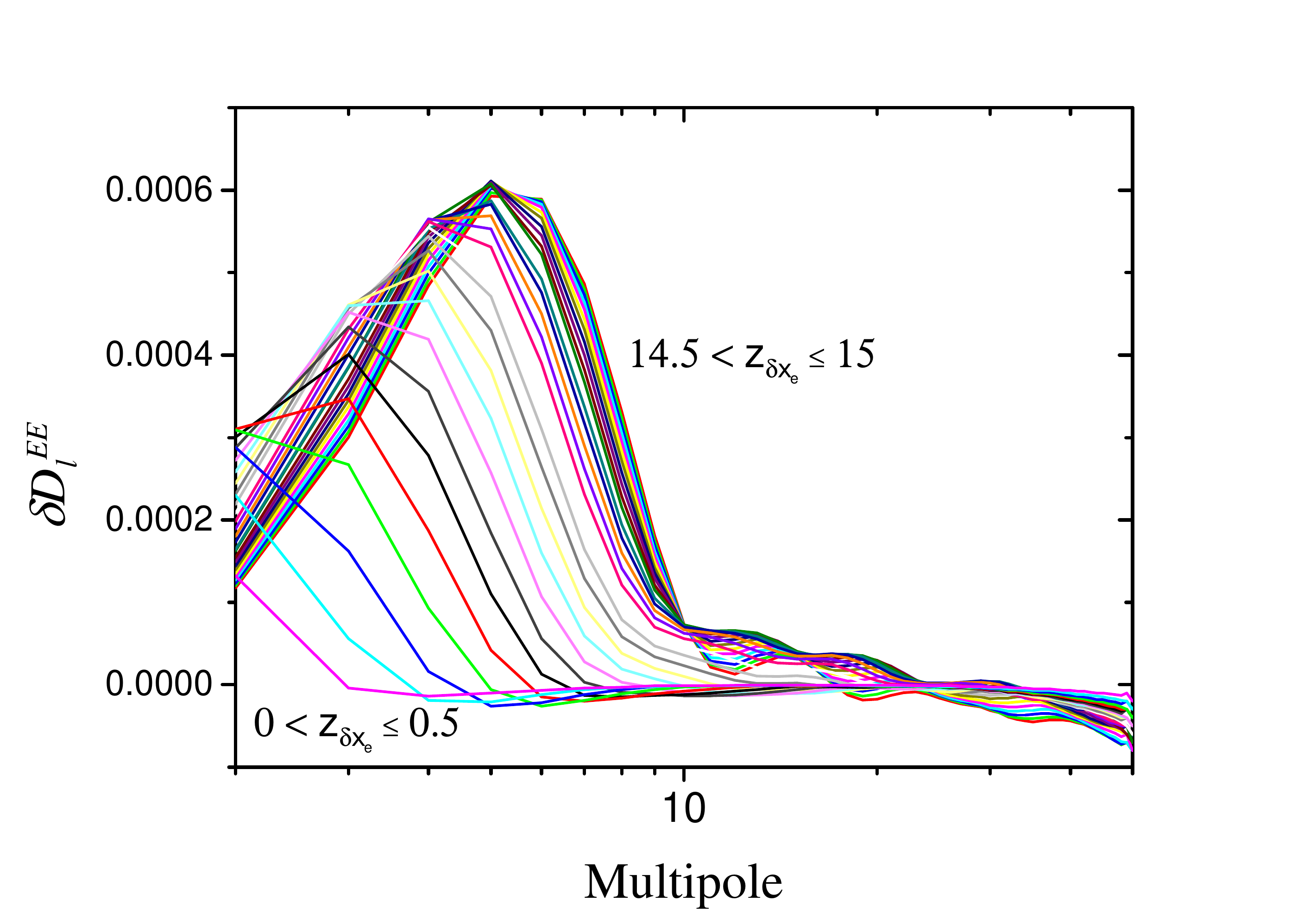}}
   \caption{Perturbed instantaneous models (left panel) and the corresponding E power spectra differences $\delta D_l^{EE}$ (right panel). The perturbations $\delta x_e$ is taken to be $0.055$, and the locations of $\delta x_e$ are taken in different redshift bins between [0,15] with $\triangle z=0.5$.}\label{figure:perturbation}
\end{figure}

In fact, even for the same $\tau$, we can have different kinds of reionization history which will lead to different power spectra, especially for E mode at small $l$, as is illustrated in Fig. \ref{figure:models}. To see the detailed effect that reionization history have on power spectra, we perturb the instantaneous model, as shown in Fig. \ref{figure:perturbation}, at different redshift bins. We choose a bin width of $\triangle z=0.5$ to segment the redshift from z=0 to 15, at each run we take the excursion value as $\delta x_e=0.055$ at only one bin while remaining the other bins $\delta x_e=0$.

The differences of power spectra $\delta D_l^{EE} = {D_l^{EE}}_{pert.} - {D_l^{EE}}_{inst.}$ are shown in right panel of Fig. \ref{figure:perturbation}. From the location of the peaks, we know that perturbations at high redshift influence relative high $\ell$ spectrum more, and low redshift perturbations have more effect at low $\ell$. We limit $\ell <50$ since the polarization power spectra at $\ell >50$ are irrelevant to our analysis of reionization.

\begin{figure}

\subfigure{\includegraphics[width=3in]{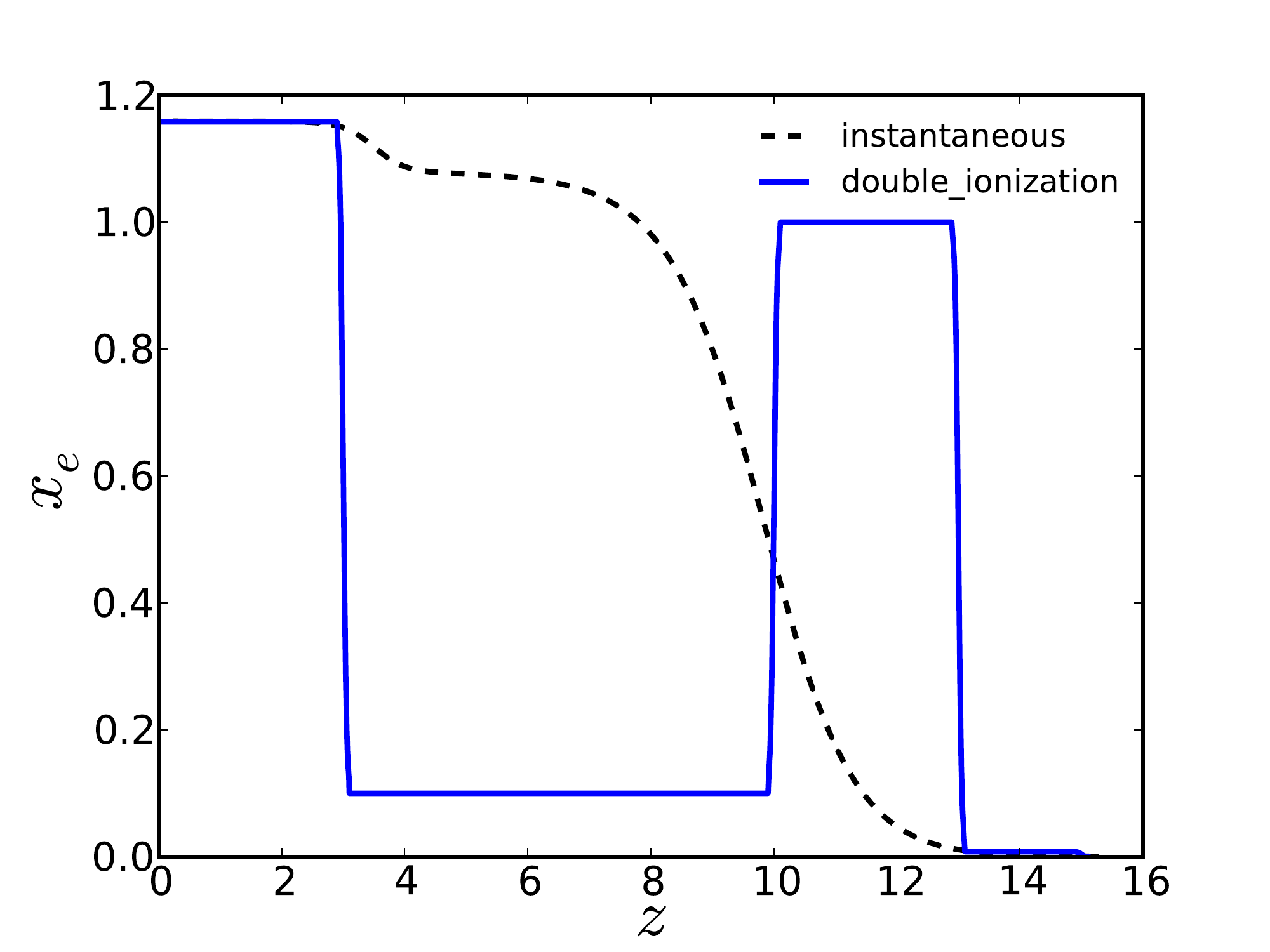}}
\subfigure{\includegraphics[width=3in]{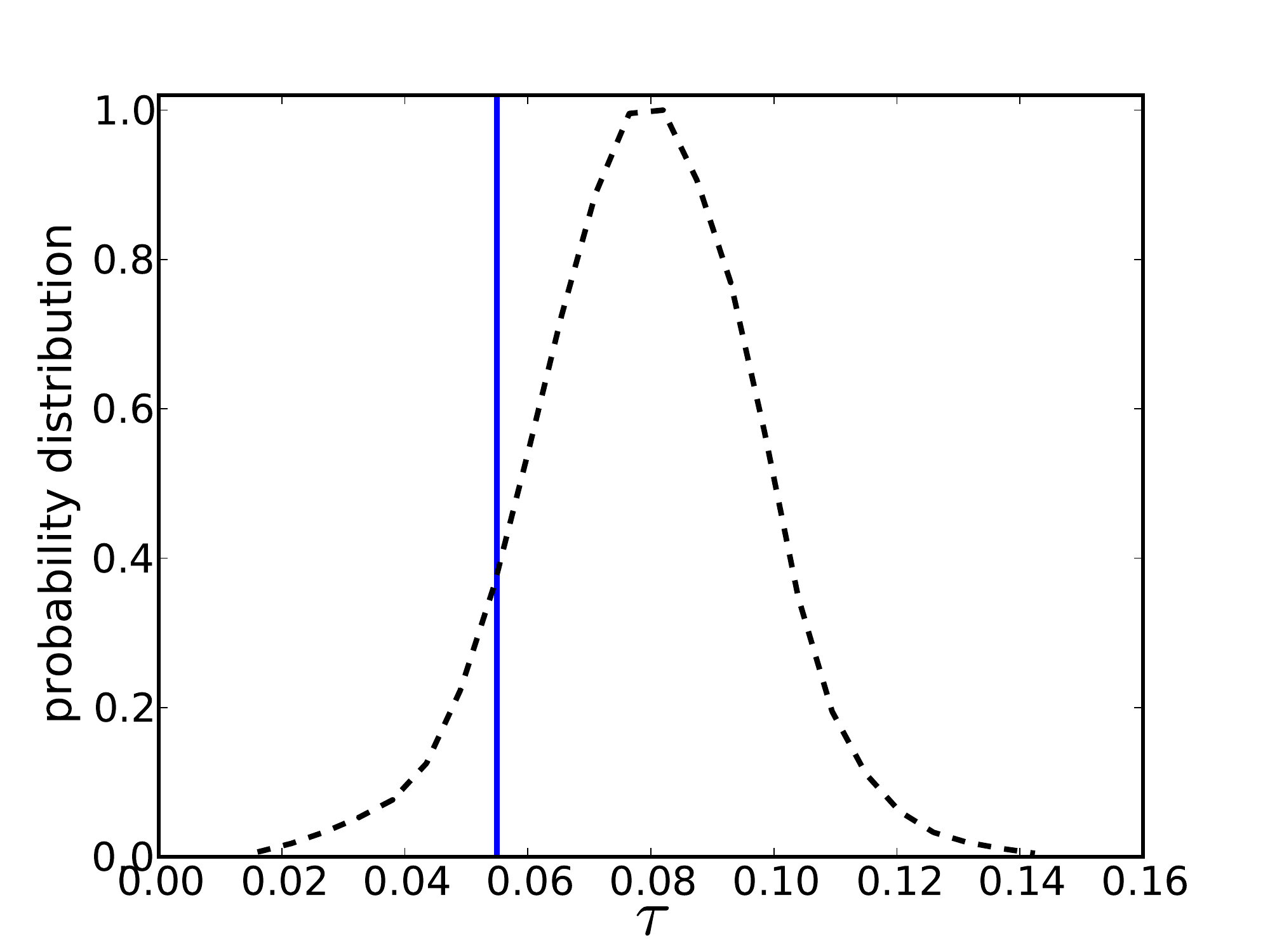}}
\caption{The left panel plots the fiducial double reionization model in solid line for simulating CMB data, and the instantaneous model derived from fitting the mocked data is shown in dashed line. The right panel shows 1 dimensional distribution constraint on $\tau$ derived from fitting the mocked data with instantaneous model, vertical line is the $\tau$ value given by the fiducial model.}\label{figure:simulation}

\end{figure}

As we know that, $\tau$ is a very important parameter when performing the data analysis with CMB, however, it will give bias when we only consider $\tau$ instead of reionization fraction $x_e(z)$. In order to see the bias introduced by instantaneous assumption,  we simulate future CMB experiments with a non-instantaneous model, and then perform the fitting with an instantaneous model, and compare the final constraints on $\tau$ with the fiducial model. The fiducial model, a double ionized $x_e$ function is plotted in left of Fig. \ref{figure:simulation} in solid line, with this model, by the integration, the optical depth is 0.055.  We simulate the CMB power spectra of a BICEPIII-like futrue CMB data with $1/100$ noise level using CAMB. Then we  perform a  global fitting analysis by using the standard $\Lambda CDM$ model with instantaneous reionization history.

The final constraint derived by the mocked data is shown in the right panel of Fig. \ref{figure:simulation}, the vertical line is $\tau=0.055$, comparing with the mean value $\tau=0.079 $, it is disfavored at about $2 \sigma C.L.$, which shows the bias from an instantaneous assumption.  Considering the importance of the reionization history, In the following, we will do a model independent analysis\cite{A. Lewis 2006} for reconstructing the reionization fraction $x_e(z)$ with the current data sets of Planck, WMAP, respectively, as well as BAO and SN.

\section{data sets employed and global fitting analysis}
Our numerical calculation of global fitting analysis are performed by using a modified CosmoMC package\cite{cosmomc} by rewritten the reionization history relevant package.

\subsection{Description of the reionization history}
We segment the epoch of reionization into several bins in redshift space, and we take the ionized fraction $x_{e}(z)$ in each bins as a constant. In this way, the reionization should not be biased by the prior assumption. The redshift bins $z_{bin}^i$, $i=1\sim 9$, are choosen as: $[0-3]$, $[3-8.5]$, $[8.5-9.5]$, $[9.5-10]$, $[10-10.3]$, $[10.3-10.6]$, $[10.6-11]$, $[11-13]$, $[13-15]$ \footnote{In principle, to do model independent analysis on reionization history, the more bins we take, the less bias for $x_e(z)$ will be priorly introduced. However, due to the limited constraining power of the current data sets, we can take $9$ bins at most. Also, we have do many optimization design for the bins, for example, adjust the range for the redshift bins, the scale of the bins and so on according to the data sets. The bins listed in the main text are the best choice form our testing.}, since from the observations we know that at $z> 15 $ our universe should be neutral and $x_e(z)$ can be neglected\cite{S. Zaroubi 2012}. In order to guarantee ${x_e}^i$ in the $i th$ bins to be a constant and the function of $x_e(z)$ to be smooth, we link ${x_e}^i$ by a $tanh$ function. In numerical calculation of the global fitting analysis, $x_e^i$  are taken as free parameters.

\subsection{Parameter space and calculations}
Our procedure are performed with the
power law $\Lambda$CDM$ + x_e(z)$ model described by the basic
parameters of $\left\{ \Omega_bh^2, \Omega_ch^2, \Theta_s, {x_e}^i,  n_s, A_s \right\}$,
where $\Omega_c h^2$ is the cold dark matter energy density parameter,
$\Omega_b h^2$ the baryon energy density
parameter, $100\Theta_s$ is the ratio (multiplied by 100)
of the sound horizon at decoupling over the angular diameter distance
to the last scattering surface, ${x_e}$ describing the reionization history and
${x_e}^i$ are the reionization fraction
parameters ${x_e}^i$ for the $ith$ redshift bin, $n_s$ and $\ln[10^{10}A_s]$
are the scalar spectral index and the primordial amplitude respectively.
During our calculation, we find that most of the background cosmological parameters, 
for example $\Omega_c h^2$, $\Omega_b h^2$, $100\Theta_s$ and $n_s$, do
not have very strong correlation with $x_e(z)$, in order to get tight constraints, 
we fixed them to the best fit values listed in table \ref{table:fiducial} derived from fitting the data with standard 6 parameters $\Lambda CDM$+instantaneous reionization model.

We free the reionization
history relevant parameters of ${x_e}^i$ in each redshift bins as well as $A_s$ which
is strongly correlated with $x_e(z)$ parameters. The top-hat priors of free parameters are
$\ln[10^{10}A_s] \in [2, 4]$ and $x_e(z) \in [0, 2.0]$.
The pivot scale is set at $k_{s0} = 0.05$ ${\rm Mpc}^{-1}$, and in the calculation
we assume an purely adiabatic initial condition.

\begin{table}
\caption {Constraints on the parameters derived from fitting current data with instantaneous model.}\label{table:fiducial}
\begin{center}
   \begin{tabular}{ c | c | c | c | c  }
   \hline			
    & ~~~WMAP+SN+BAO~~~ & ~~~sddev~~~ & ~~~Planck+SN+BAO~~~ & ~~~sddev~~~ \\
   \hline
  $\Omega_b h^2$ & 0.02246 & 0.00043 & 0.02228 & 0.00014\\
  $\Omega_c h^2$ &  0.1170 & 0.0021 & 0.1192 & 0.0011\\
  $100\theta_{MC}$ & 1.03948 & 0.00210 & 1.04084 & 0.00030\\
  $\tau$ & 0.085 & 0.013 & 0.082 & 0.017\\
  ${\rm{ln}}(10^{10} A_s)$ & 3.094 &0.029 & 3.097 & 0.033\\
  $n_s$ & 0.9667 & 0.0101 & 0.9641 & 0.0040\\

  \hline
  \end{tabular}
\end{center}
\end {table}

\subsection{Current Observational Data}

In our analysis, we consider the following cosmological probes: i)
power spectra of CMB temperature and polarization anisotropies
released by WMAP 9 years and Planck2015 data;
ii) the baryon acoustic oscillation in the galaxy
power spectra; iii) luminosity distances of type Ia supernovae.

For the Planck data from the 2015-year data release
\cite{Planck2015b}, we use the low-$\ell$ temperature-polarization likelihood at multipoles $2\leq \ell \leq 29$($low TEB$) and high-$\ell$ likelihood combining TT,TE,and EE power spectra at multipoles $\ell \geq 30$($PlikTT,EE,TE$), we will call the whole data used as Planck for short.  Also, we have considered
the $9$ year WMAP temperature and  polarization spectra\cite{WMAP9} which are provided
by CosmoMC package, it can be called as WMAP for short.

Baryon Acoustic Oscillations provides an efficient method for
measuring the expansion history by using features in the
clustering of galaxies within large scale surveys as a ruler with
which to measure the distance-redshift relation\cite{bao}. Since the current BAO data are
not accurate enough for extracting the information of $D_A(z)$ and
$H(z)$ separately \cite{okumura}, one can only determine an
effective distance \cite{baosdss}:
\begin{equation}
D_V(z)=[(1+z)^2D_A^2(z)cz/H(z)]^{1/3}~.
\end{equation}
Following the Planck analysis\cite{Planck2015b}, in this paper we
use the BAO measurement from the 6dF Galaxy Redshift Survey
(6dFGRS) at a low redshift ($r_s/D_V (z = 0.106) = 0.336\pm0.015$)
\cite{6dfgrs}, and the measurement of the BAO scale based on a
re-analysis of the Main Galaxy Sample(MGS) from Sloan
Digital Sky Survey (SDSS) Data Release 7($D_V/r_s (z=0.15) =4.466 \pm 0.168$)\cite{A.J. Ross}
, BAO signal from BOSS DR11 LowZ($D_V / r_s (0.32) = 8.250 \pm 0.170$)\cite{sdssdr11}and the
BAO signal from BOSS CMASS DR11 data at ($D_V/r_s(0.57)=13.773 \pm 0.134$) \cite{sdssdr11}.

Finally, we include data from Type Ia supernovae, which consists
of luminosity distance measurements as a function of redshift,
$D_L(z)$. In this paper we use the supernovae data set,
``Union2.1'' compilation, which includes 580 high-redshift Type Ia supernovae
reprocessed by Ref.\cite{Union2.1}. When calculating the
likelihood, we marginalize the nuisance parameters, like the
absolute magnitude $M$.

\subsection{Principle component analysis method}

By doing the global fitting, we get constraints on $x_e(z)$ for each bins. The constraints of $x_e^i$ are correlated and, usually, it is considered that the correlated constraints are not the physical solutions. Basing on the correlated constraints and the associate correlation covariance, one can construct a basis of $x_e^i$, with which we can get uncorrelated constraints on ionized fraction parameters by adopting the PCA method \cite{Huterer_LPCA, Huterer_PCA}.

The covariance matrix of $x_e^i$s can be derived from the MCMC fitting procedure. In practice, we perform PCA method with $\delta x_e^i = x_e^i - {x_e}_{\rm inst.}^i$ instead, where ${x_e}_{\rm inst.}$ is instantaneous function obtained from the best fit result from the same data, assuming that the deviation from intaneous model, $\delta x_e^i$,  could be treated as fluctuations.  One can simply prove that the covariance matrices of $x_e^i$ and $\delta x_e^i$ are the same, reads:
\begin{equation}
C=<(x_e^i-\langle x_e^i\rangle)(x_e^j-\langle x_e^j\rangle)^T>=<(\delta x_e^i-\langle \delta x_e^i\rangle)(\delta x_e^j-\langle \delta x_e^j\rangle)^T>=\langle \vec{p}\vec{p}^T\rangle-\langle \vec{p}\rangle \langle \vec{p}^T\rangle,
\end{equation}
$\vec{p}$ is the vector of $\delta x_e^i$ parameters and $\vec{p}^T$ is its transpose, and the Fisher matrix of $\vec{p}$ is $F=C^{-1}$. In order to get uncorrelated information of $\delta x_e^i$, we should rotate $\vec{p}$ into a basis where the covariance matrix (or the Fisher matrix) is diagonal. To do that, we rotate the Fisher matrix by an orthogonal matrix $W$,
\begin{equation}
F=W^TDW,
\end{equation}
where $D$ is diagonal. The new parameters for ionized fraction parameters can now be written as $\vec{q}=W\vec{p}$ which are uncorrelated with each other because they have the diagonal covariance matrix $D^{-1}$. The $q_i$ are supposed to be the principal components (PCs) and the rows of the decorrelation matrix W are the window functions (or weights) which define the relations between the original parameters and the uncorrelated parameters $q_i$.

There are many matrices that can diagonalize $F$. The special type of decorrelation matrix which absorbs the diagonal elements of $D^{\frac{1}{2}}$ into the rows of W mentioned above, multiplying any orthogonal matrix $O$, $W^{*}=OD^{\frac{1}{2}}W$, can also diagonalize $F$ and make the parameters $q$ uncorrelated. In order to get uncorrelated $q_i$ which are physical without artificial treatment, we choose to adopt the following two kinds of realization:

I. Normal principal component analysis: diagonalizing $F$ by an orthogonal matrix $W$, and then we order the eigenvalues of the diagonal matrix from small to large, by doing this we can fix the form of the orthogonal $W$ matrix. In this case, we can filter out the better constrained modes as well as the noisy modes. With the better constrained modes, we can reconstruct $x_e(z)$, which should be better constrained.

II. We do the local PCA by choosing the decorrelation matrix $\widetilde{W}=F^{1/2}\equiv W^{T}D^{\frac{1}{2}}W$, and normalize $\widetilde{W}$ by making its rows sum to unity, which can ensure $q(z)$ standing for instantaneous model, which means $\delta x_e(z) = 0$. This choice has the advantage that the weights of $x_e^i$ are almost positive defined and fairly well localized in the redshift bins.

\section{Numerical results}

\subsection{Constraints from global fitting analysis}

\begin{figure}
   \subfigure{\includegraphics[width=3in]{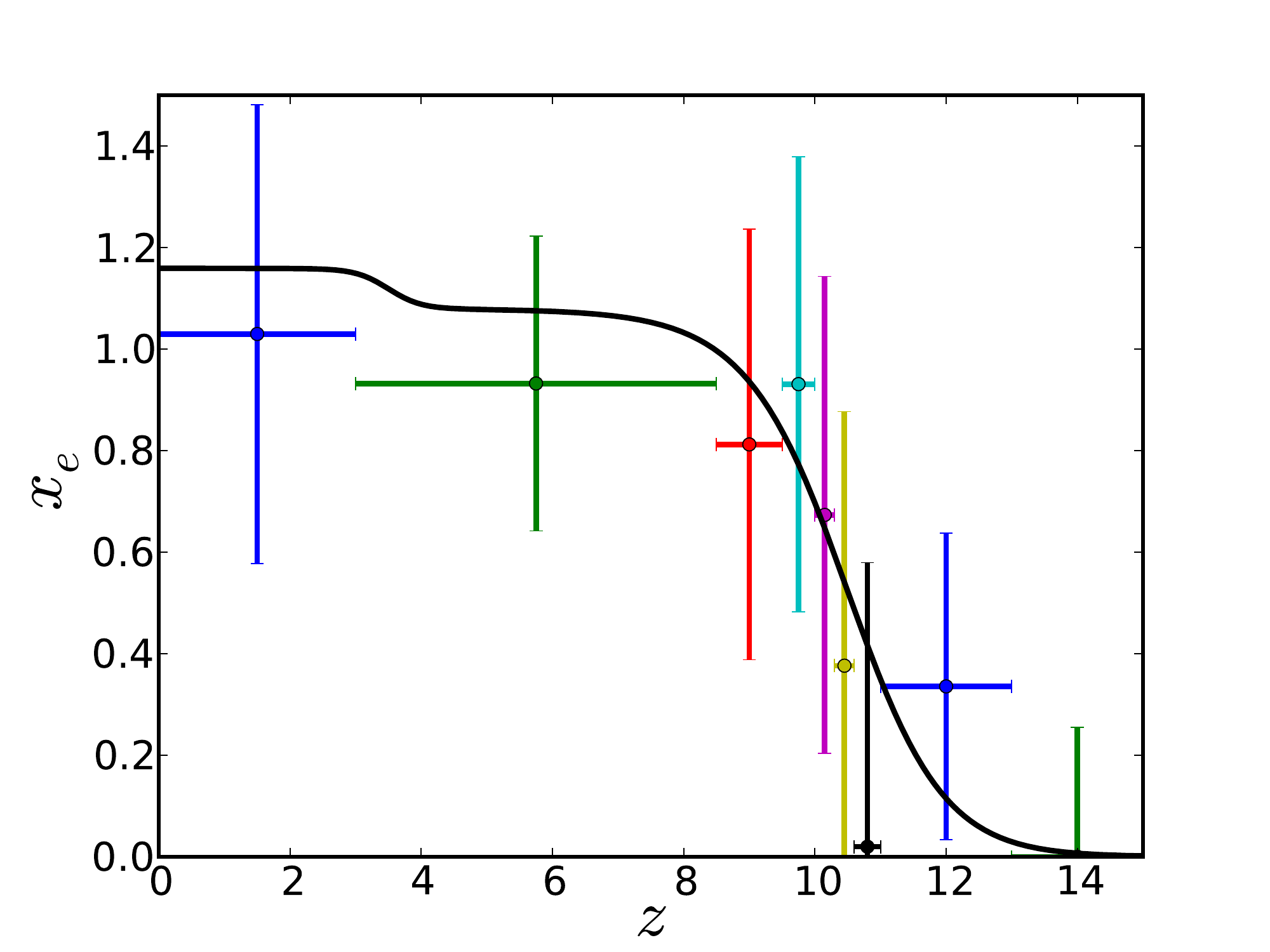}}
   \subfigure{\includegraphics[width=3in]{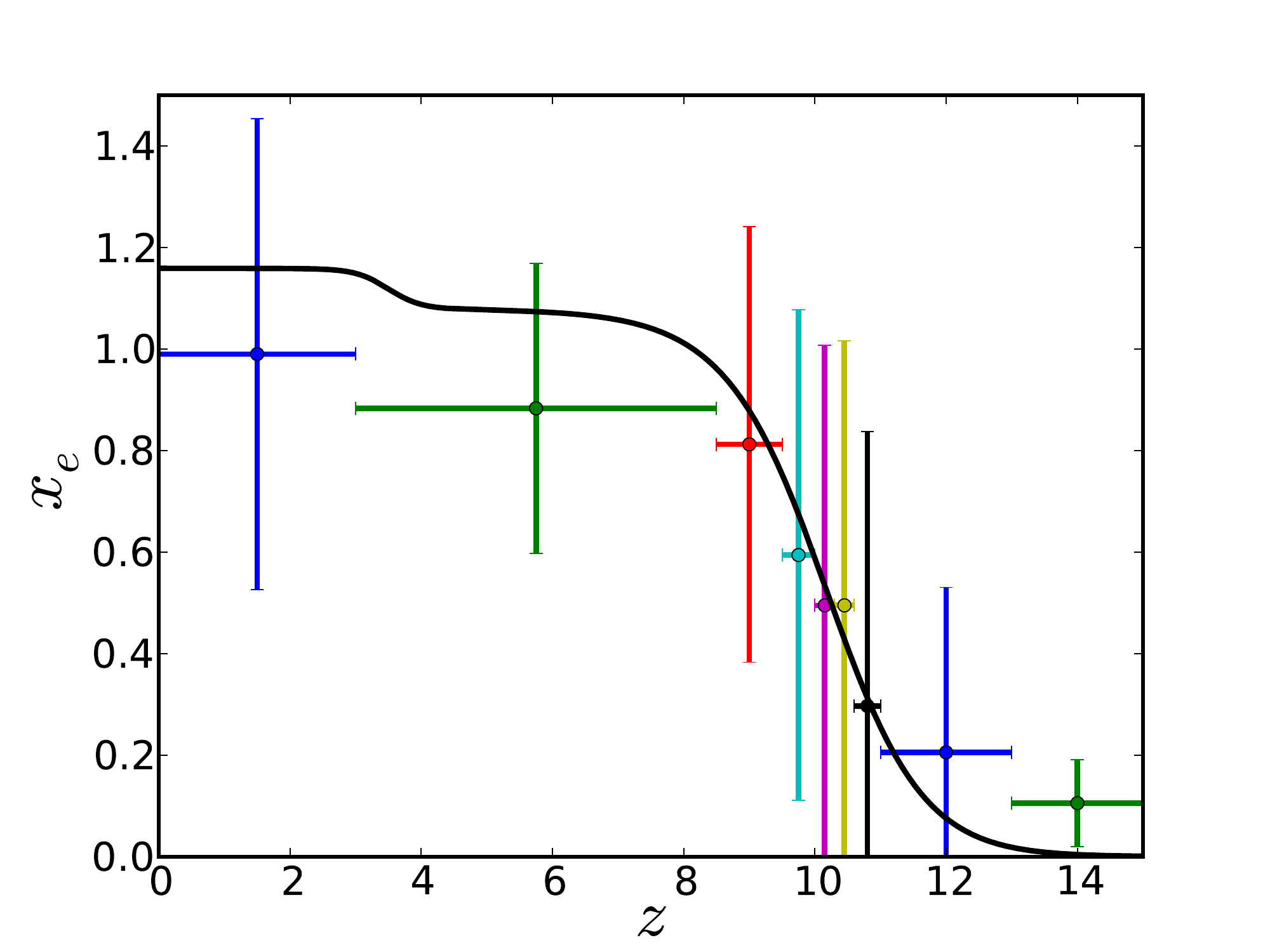}}

   \caption{Constraints on reionization bins from WMAP+SN+BAO data(left) and Planck+SN+BAO data(right), points show best fit values, vertical error bars are the 1$\sigma$ errors, and the horizontal bars show the width of the redshift bins.}\label{figure:comparison_wmap}
\end{figure}

In Fig. \ref{figure:comparison_wmap} we present the $1\sigma$ constraints on the binned redshift reionization model by using the data combination of WMAP+SN+BAO (left panel) and Planck+SN+BAO (right panel), respectively. In order to make comparison, we also perform global fitting with instantaneous model.

From the results of fitting with the data combination of WMAP+SN+BAO, the best fit values of the bins manifest a tendency that the reionization should last for a period of time to realize that it is totally ionized today and $x_e(z)\sim 0$ at higher redshift in $z\subset [13, 15]$. Also, we plot the best fit value of the instantaneous model in the figure by a black solid line, it is consistent with the binned models at about $1\sigma$ $C.L.$. However, there are a few bins present featured structure which deviate from an instantaneous behavior, for example, the best fit value of the 7th bin is much lower than the instantaneous model while $8th$ bins are much higher, and the deviations are at about $1\sigma$. From detail calculation, we find that in the small $l$ region, the power spectrum obtained from bin model is smaller than instantaneous one, on large $l$ region they are very same to each other, and the binned model fits WMAP+SN+BAO data better, since it can produce much lower power on large scale. We also make comparison of the constraints on $\tau$ between the two models. From the binned redshift model we get $\tau=0.088\pm0.012$ and  $\tau=0.085\pm0.013$ derived from the instantaneous model, and they are consistent with each other, which show that with the current data sets, an instantaneous model do not bias the constraints on constraining $\tau$, since the current data sets are not accurate enough to distinguish the two scenarios.

The detailed numerical constraints on cosmological parameters are listed in table \ref{table:fiducial} and  table \ref{table:fitting} for the instantaneous model and the binned model respectively.

Comparing with WMAP, the constraints from Planck is a little bit weaker, which can also be seen from the table \ref{table:fitting}. The main reason is that the error estimates for WMAP data do not reflect the true uncertainty in foreground removal, the WMAP do not know the actual dust components\cite{Planck2015a}. In right panel of Fig. \ref{figure:comparison_wmap}, we plot the constraints from Planck + BAO+ SN. These two models are consistent with each other at about $1\sigma~C.L.$, except the last bin. There is an obvious deviation from the instantaneous model result, which still support that there maybe a bump at the beginning of reionization. We also find that the binned model give much lower power in large scale comparing with the instantaneous model. The combination of Planck data give the optical depth as $\tau=0.080 \pm 0.013$ and the primordial amplitude as $\ln[10^{10}A_s]=3.094 \pm 0.025$, the errors are slightly smaller than the instantaneous model.

\begin{table}
\caption {Constraints on $x_e^i$, $\tau_{\rm{derived}}$ and $\ln[10^{10}A_s]$.}\label{table:fitting}
\begin{center}
   \begin{tabular}{ c | c | c | c | c  }
   \hline			
    & ~~~WMAP+SN+BAO~~~ & ~~~sddev~~~ & ~~~Planck+SN+BAO~~~ & ~~~sddev~~~ \\
   \hline
  $x_e^1$ & 1.02972 & 0.45188 & 0.99011 & 0.46380\\
  $x_e^2$ & 0.93224 & 0.29041 & 0.88331 & 0.28553\\
  $x_e^3$ & 0.81219 & 0.42421 & 0.81236 & 0.42931\\
  $x_e^4$ & 0.93097 & 0.44809 & 0.59427 & 0.48329\\
  $x_e^5$ & 0.67332 & 0.46978 & 0.49518 & 0.51254\\
  $x_e^6$ & 0.37639 & 0.50048 & 0.49521 & 0.52113\\
  $x_e^7$ & 0.01988 & 0.55960 & 0.29709 & 0.54006\\
  $x_e^8$ & 0.33563 & 0.30148 & 0.20571 & 0.32496\\
  $x_e^9$ & 0.00001 & 0.25545 & 0.10564 & 0.08578\\
  $\tau_{\rm{derived}}$ & 0.088 & 0.012 & 0.080 & 0.013\\
  $\ln[10^{10}A_s]$ & 3.098 & 0.023 & 3.094 & 0.025\\

  \hline
  \end{tabular}
\end{center}
\end {table}

\subsection{Constraints on ionization fraction parameters by adopting normal PCA method}
PCA method would provide tighter constraints on the model parameters, since it performs an estimation on the noise of each mode from the data fitting analysis. By ignoring the noisy modes, it provides a useful way for measuring cosmological parameters. In the following, we will adopt PCA to tighten the constraints.

For normal PCA, we should diagonalize $F$ by a matrix $W$, and realize $F=W^TDW$. The diagonal elements of $D$ are $d_i$, and each of uncorrelated parameters $q_i$ has an error $\sigma(q_i)=d_i^{-1/2}$. We order $d_i$ so that $\sigma(q_1)<\sigma(q_2)<....<\sigma(q_N)$. There are many orthogonal matrixes that can diagonal F, however, after ordering the diagonal element of $D$, the decorrelation matrix $W$ should be fixed. The rows of $W$ are the eigenvectors $e_i(z)$.

From the errors of the eigenvectors, we can judge which modes are better constraints. By keeping those good modes, we can reconstruct $x_e^i$ as:
\begin{equation}\label{eq7}
x_e^i(M)=\sum_{j=1}^{M} q_{j} e_j^i+{x_e}_{\rm inst.}^i,
\end{equation}
where $M$ stands for the number of modes, i and j are the bin order and the mode order respectively.

\begin{figure}
\centering
   \includegraphics[width=0.8\textwidth]{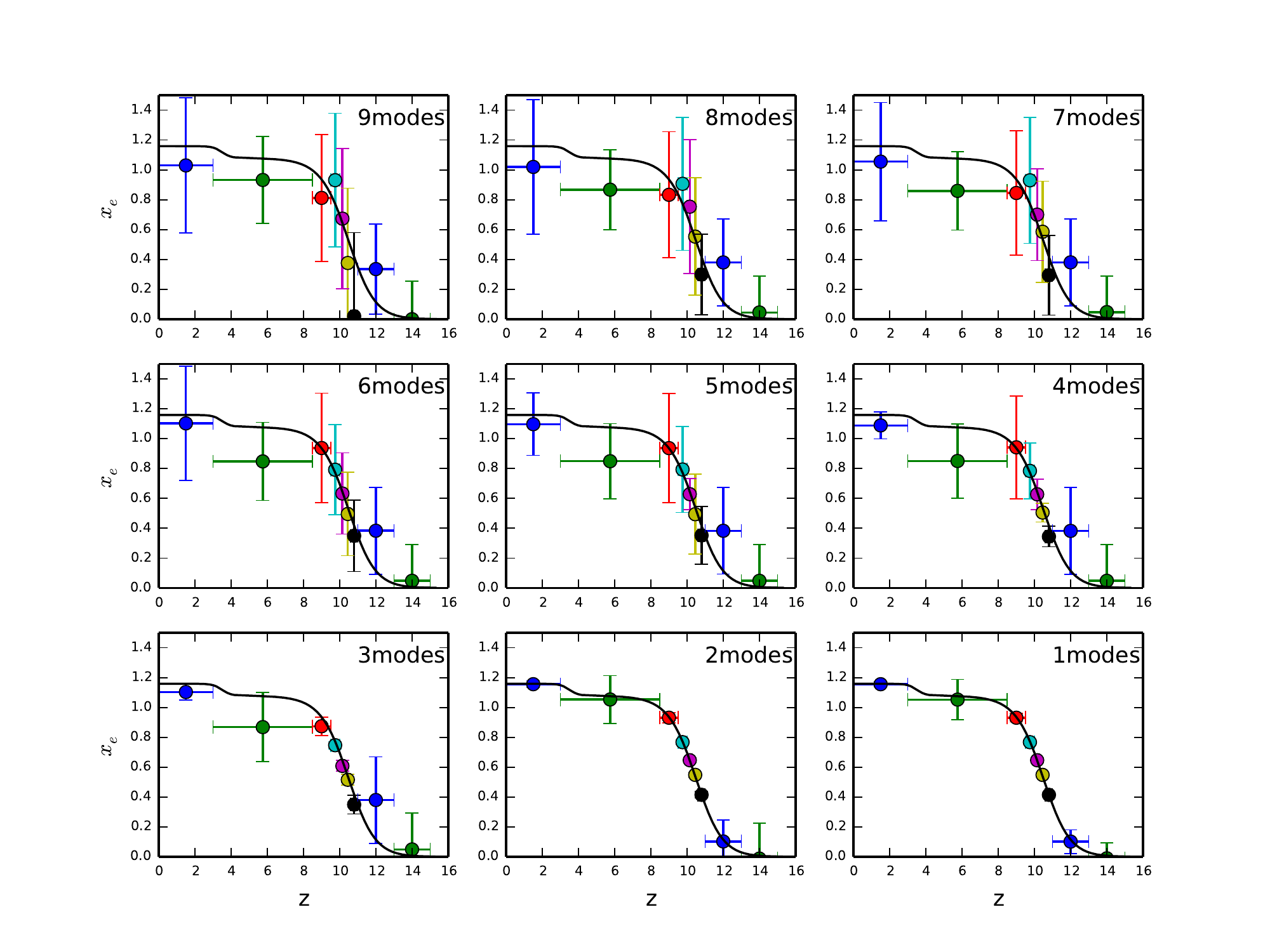}
   \caption{The reconstructed $x_e(z)$ evolution using PCA by fitting with data sets of WMAP+SN+BAO.}\label{figure:wmappca}
\end{figure}

\begin{figure}
\centering
   \includegraphics[width=0.8\textwidth]{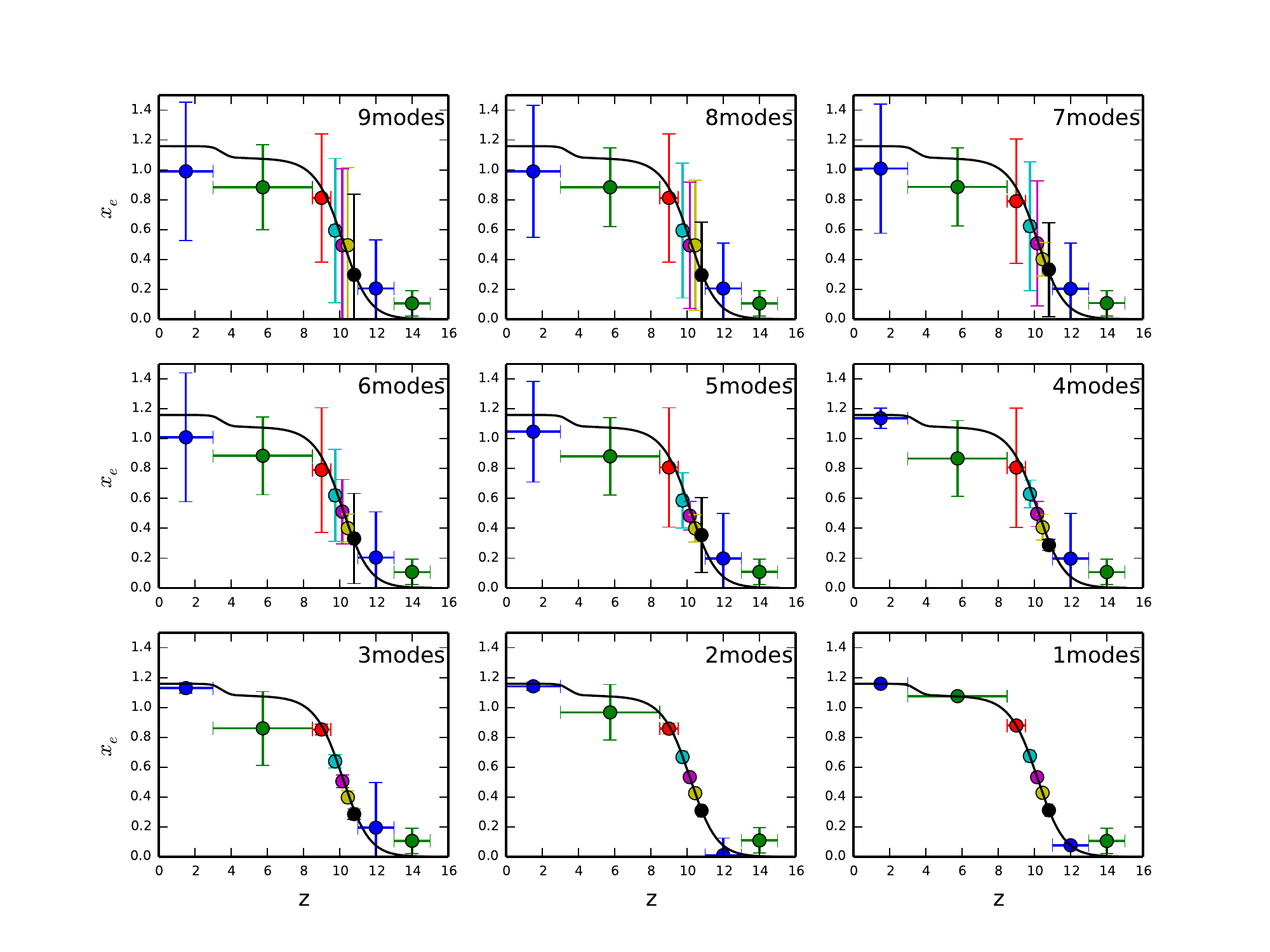}
   \caption{The reconstructed $x_e(z)$ evolution using PCA by fitting with data sets of Planck+SN+BAO.}\label{figure:planckpca}
\end{figure}

\begin{figure}

\subfigure{\includegraphics[width=3in]{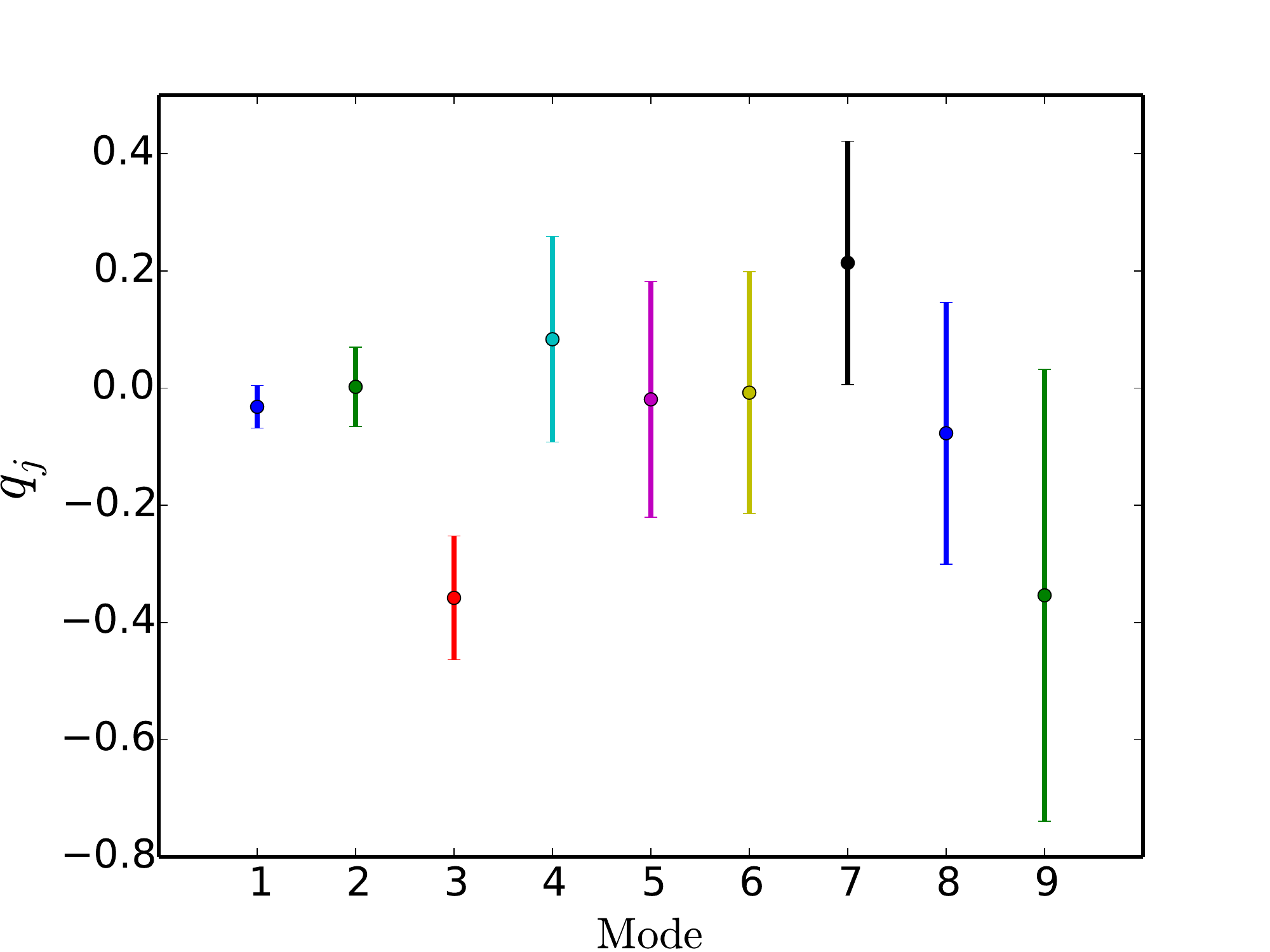}}
\subfigure{\includegraphics[width=3in]{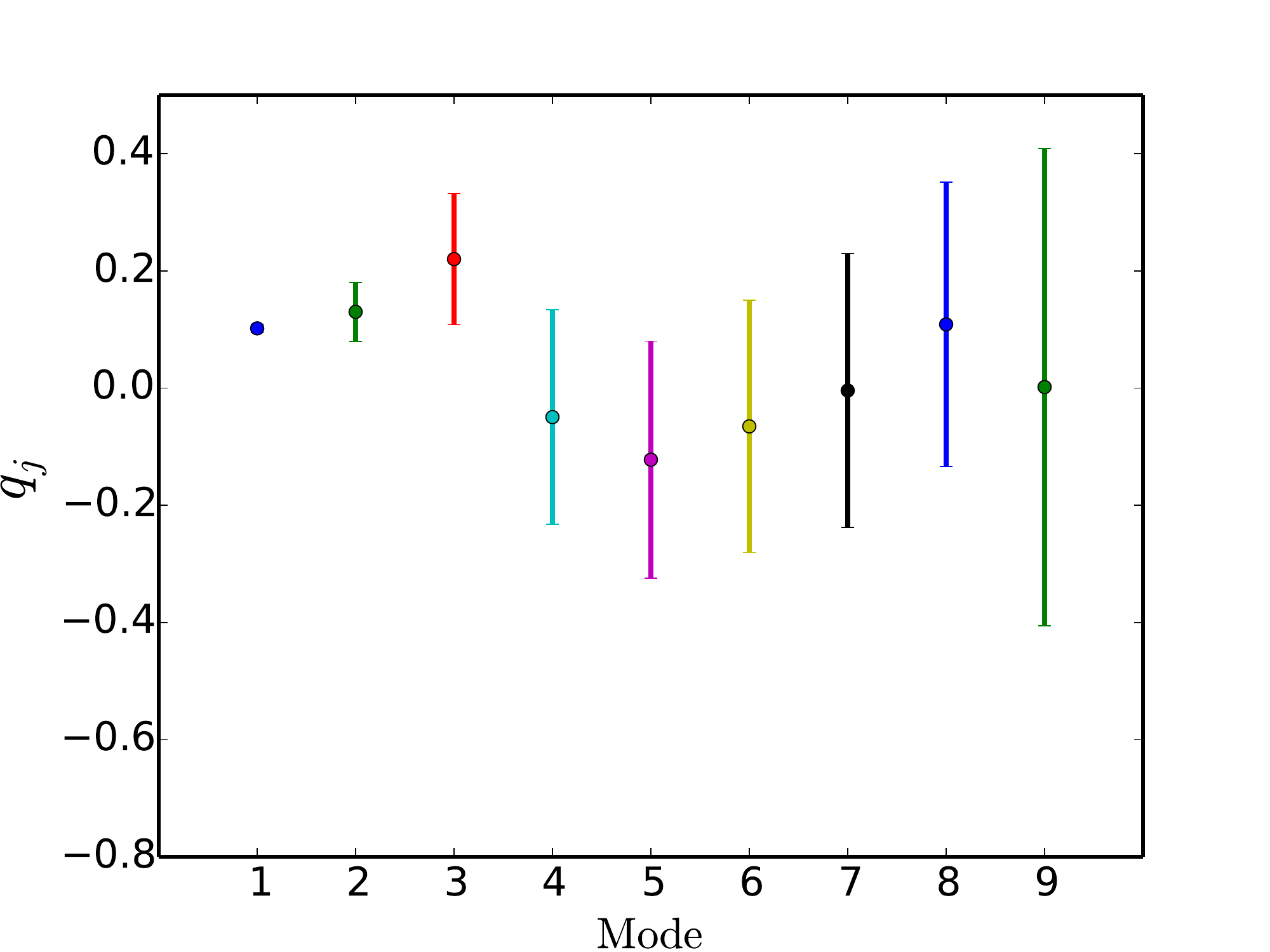}}
\caption{The uncorrelated parameters $q_j$s(PCA) and their $1\sigma$ errors for WMAP+SN+BAO(left) and Planck+SN+BAO(right)}\label{figure:alphavalue}
\end{figure}

We plot the uncorrelated parameters $q_j$ in Fig. \ref{figure:alphavalue}, and with these components, we reconstruct the behavior of ionized fraction parameters. In order to get better constraints on $x_e^i$, we truncate the badly constrained modes and just consider the contribution of the good ones. With Eq.(6), we compare several cases for adopting different number of modes respectively in Fig. \ref{figure:wmappca} for the data combination of WMAP and Planck, respectively. We plot the reconstructed $x_e(z)$ by adopting different number of modes. With all the modes, the 9 bins, we will get the same $x_e(z)$ function as shown in Fig. \ref{figure:comparison_wmap} in previous section. When we ignore the most noisy mode and adopt the first 8 modes, we can get a reconstructed $x_e(z)$ in the second plot, in which we see that the featured structure becomes more obvious, for example in the results given by fitting with WMAP+BAO+SN, the deviation in 2nd and 8th bin increase to $1\sigma~ C.L.$ region. With ignoring more modes, the variance becomes smaller while the bias becomes greater.

In fact, reconstructing $x_e(z)$, two things should be balanced properly, a). Adopting only few modes into the reconstruction in order to avoid too much noise might lead to obvious bias, since when you abandon more noisy modes, at the same time you also lose information from the data sets which can lead to bias.  b). Adopting more modes to avoid the risk of deviating from the original $x_e(z)$, at the same time it brings too much noise which will weaken the final constraints on reionization process. To quantify such balance, we do the estimation of the so called $risk$ following the paper \cite{Huterer_PCA}, where

\begin{equation}\begin{split}
risk & =bias^2 + variance \\
     & =\sum_{i=1}^{N} [x_e^i(M)-{\bar{x_e^i}}]^2+\sum_{i=1}^{N}{\sigma}^2(x_e^i(M))
\end{split}\end{equation}

\begin{figure}

\subfigure{\includegraphics[width=3in]{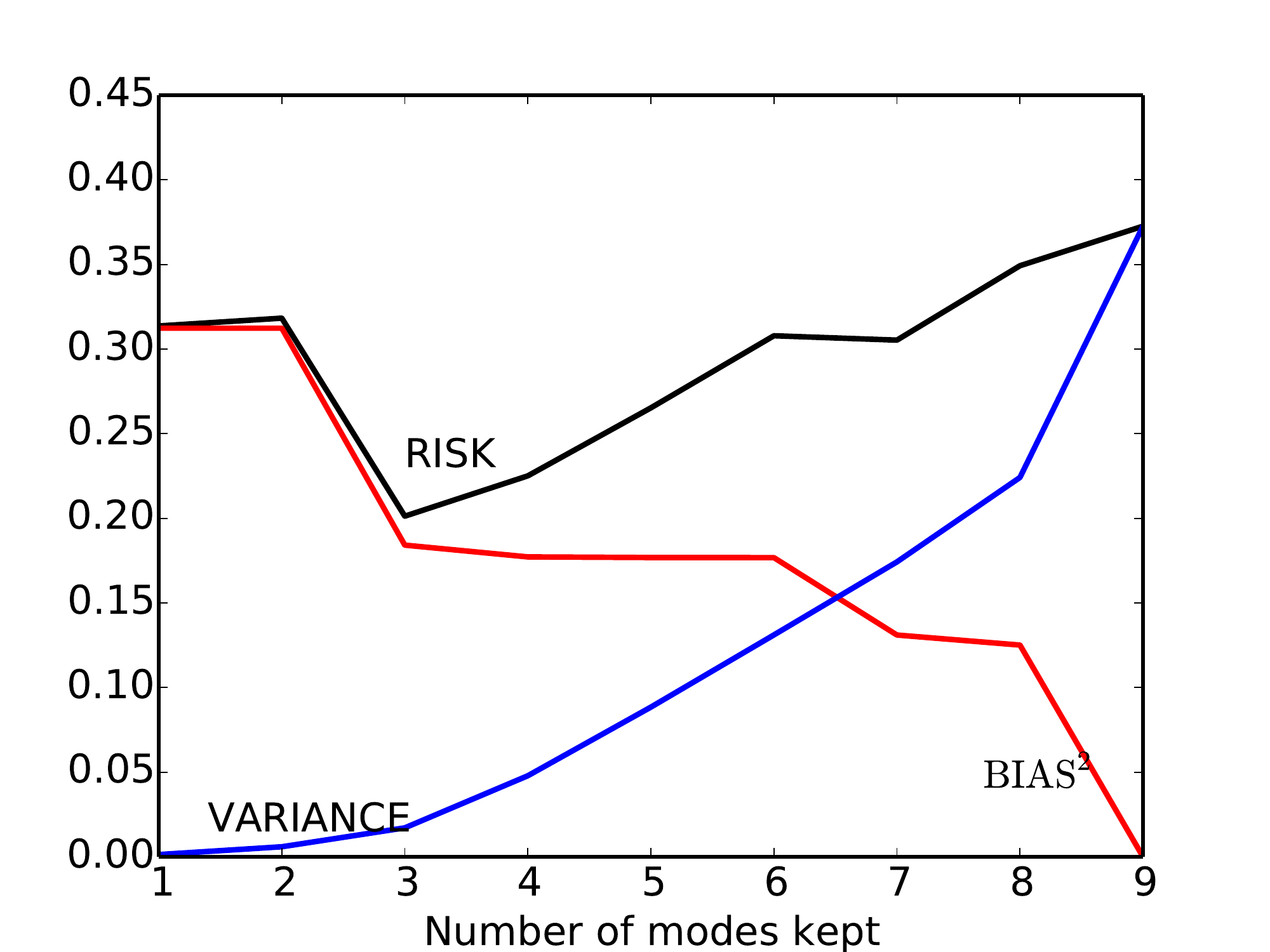}}
\subfigure{\includegraphics[width=3in]{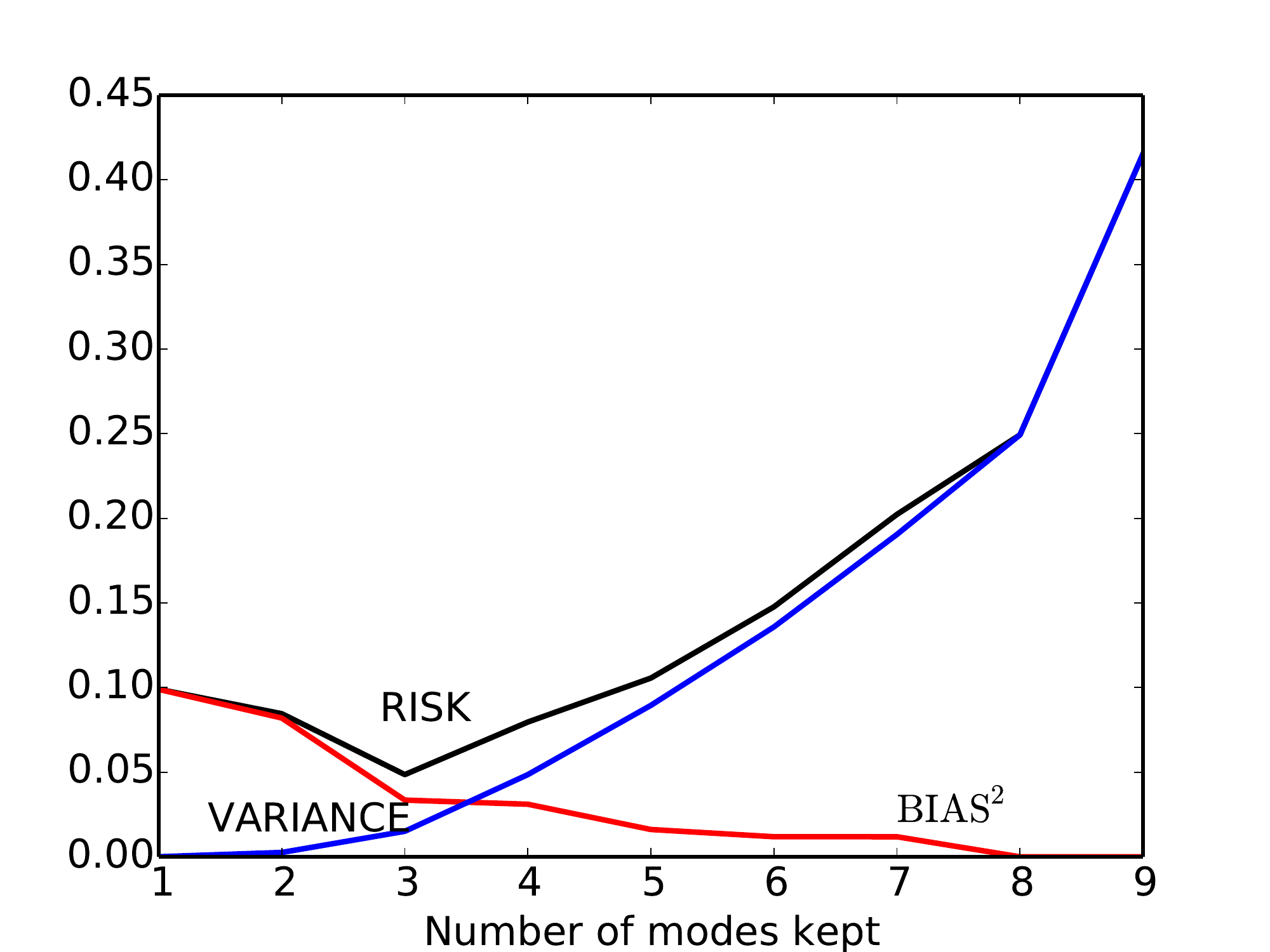}}
\caption{Illustration of the minimization of risk for WMAP+SN+BAO(left) and Planck+SN+BAO(right)}\label{figure:risk}
\end{figure}

Here, $x_e^i(M)$ stands for the value of reconstructed $x_e^i$ by taking into account M modes at the redshift $z_i$, and $\sigma(x_e^i)$ is its corresponding uncertainties. The $\bar{x_e^i}$ denotes the original value of $x_e^i$. $N$ denotes the total number of bins. Thus, by using the Eq.(7), the $risk$ can be regarded as the function about number of modes to be kept, $M$. In the Fig. \ref{figure:risk}, we illustrate the $risk$ value of considering different number of modes. Obviously, $M=3$ can give the minimal risk, thus keeping the first 3 modes is the best choice to minimize the $risk$. With first three modes, the reconstructed $x_e(z)$ are shown in the upper right panel of Fig. \ref{figure:wmappca}. Based on this result, we find that most center values of the bins can be consistent with a instantaneous ionized model. However, there are still three bins that show slight deviation from it, such as the 1st, 2nd and 7th bins, the deviation is at about $1 \sigma ~C.L.$. The $x_i$ in the redshift $0<z<3$ behaves smaller than the value of instantaneous value, the $x_e^i$ in the redshift $3<z<8.5$ is little smaller than instantaneous model value. While in redshift $11<z<13$ tends greater than instantaneous value, which imply maybe the resolution of reionization cannot be simply characterized by a monotonic function. We also use the recently released Planck data to do the same analysis, almost get the similar result as we can see from the Fig. \ref{figure:planckpca}. One interesting thing for Planck is that there is a very strange deviation at the last bin, and it exists at all cases, which means this deviation is the most useful information, which we need pay more attention to it.

\subsection{Constraints on ionization fraction parameters by local PCA}
Absorbing the diagonal elements of $D^{1/2}$ into orthogonal $W$, and multiplying an orthogonal matrix, is another useful realization, where we adopt $\widetilde{W}\equiv W^{T}D^{1/2}W$, as the decorrelation matrix, then the uncorrelated parameters can be written as $\vec q = \widetilde{W} \vec p$\cite{Huterer_LPCA}. The advantage of this choice is that the modes are localized distributed in each redshift bin and the weight of each mode is almost positive This kind of choice is considered as a useful basis for achieving the uncorrelated quantities, and is widely used in the analysis of the uncorrelated galaxies power spectrum\cite{M. Tegmark} as well as the equation of state parameters of dark energy\cite{{Huterer_LPCA},{G.B. Zhao2008},{G.B. Zhao2010},{W. Zheng}}.

\begin{figure}

\subfigure{\includegraphics[width=2.8in]{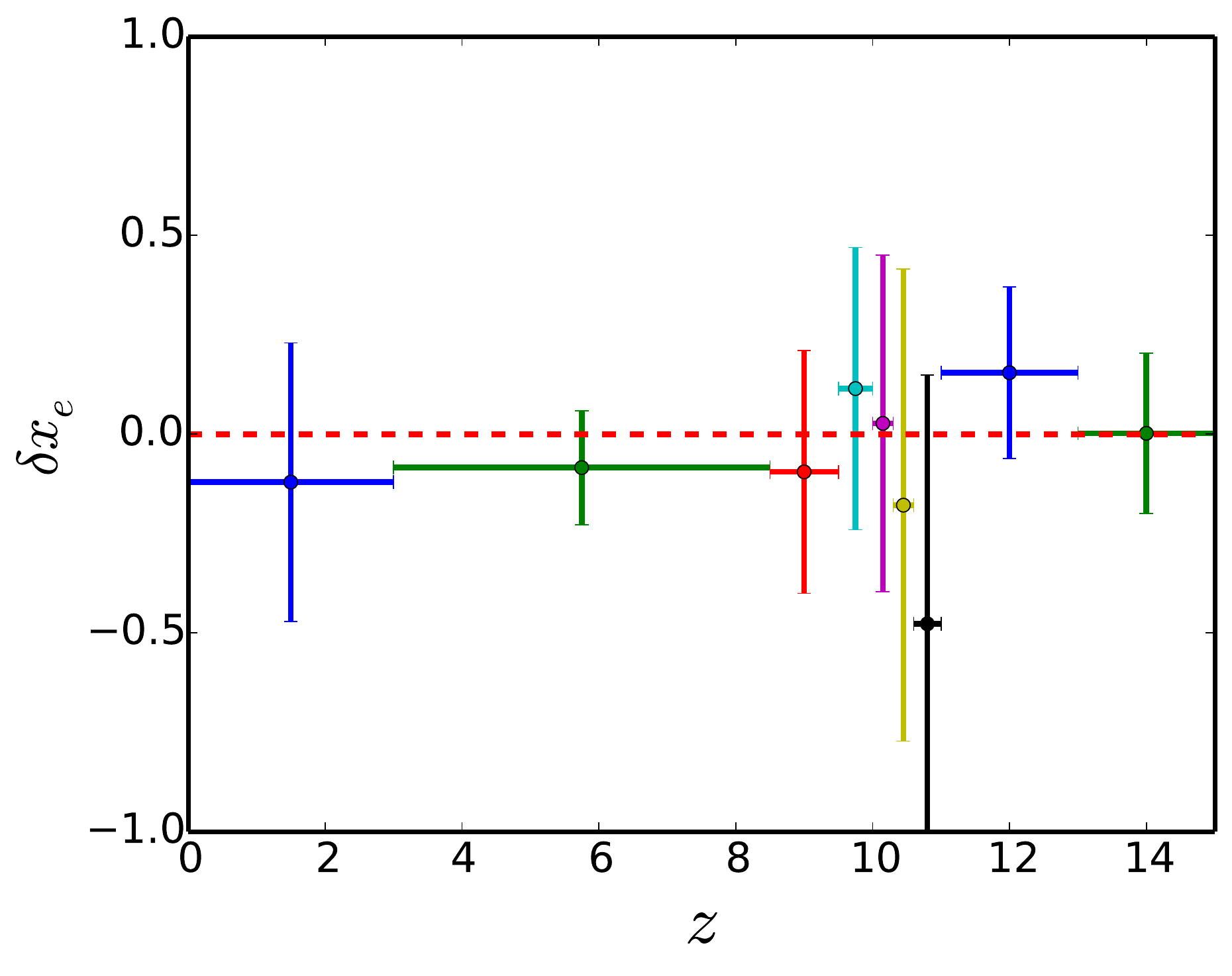}}
\subfigure{\includegraphics[width=2.8in]{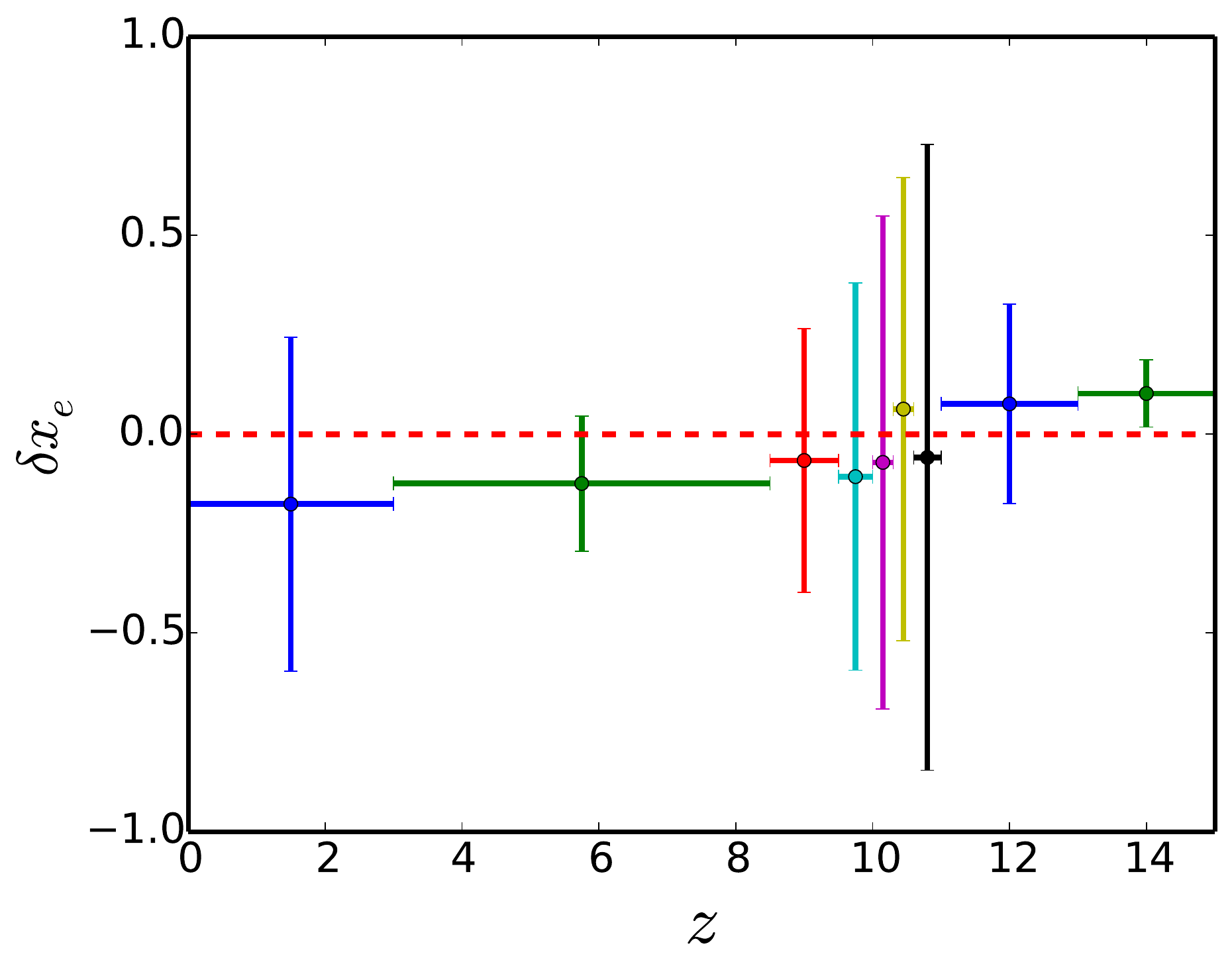}}
\caption{Uncorrelated band-power estimates of reduced reionization fraction $\delta x_e(z)$(LPCA) for WMAP+SN+BAO(left) and Planck+SN+BAO(right), vertical error bars show the $1\sigma $ error bars}\label{figure:qvalue}
\end{figure}

\begin{figure}

\subfigure{\includegraphics[width=3in]{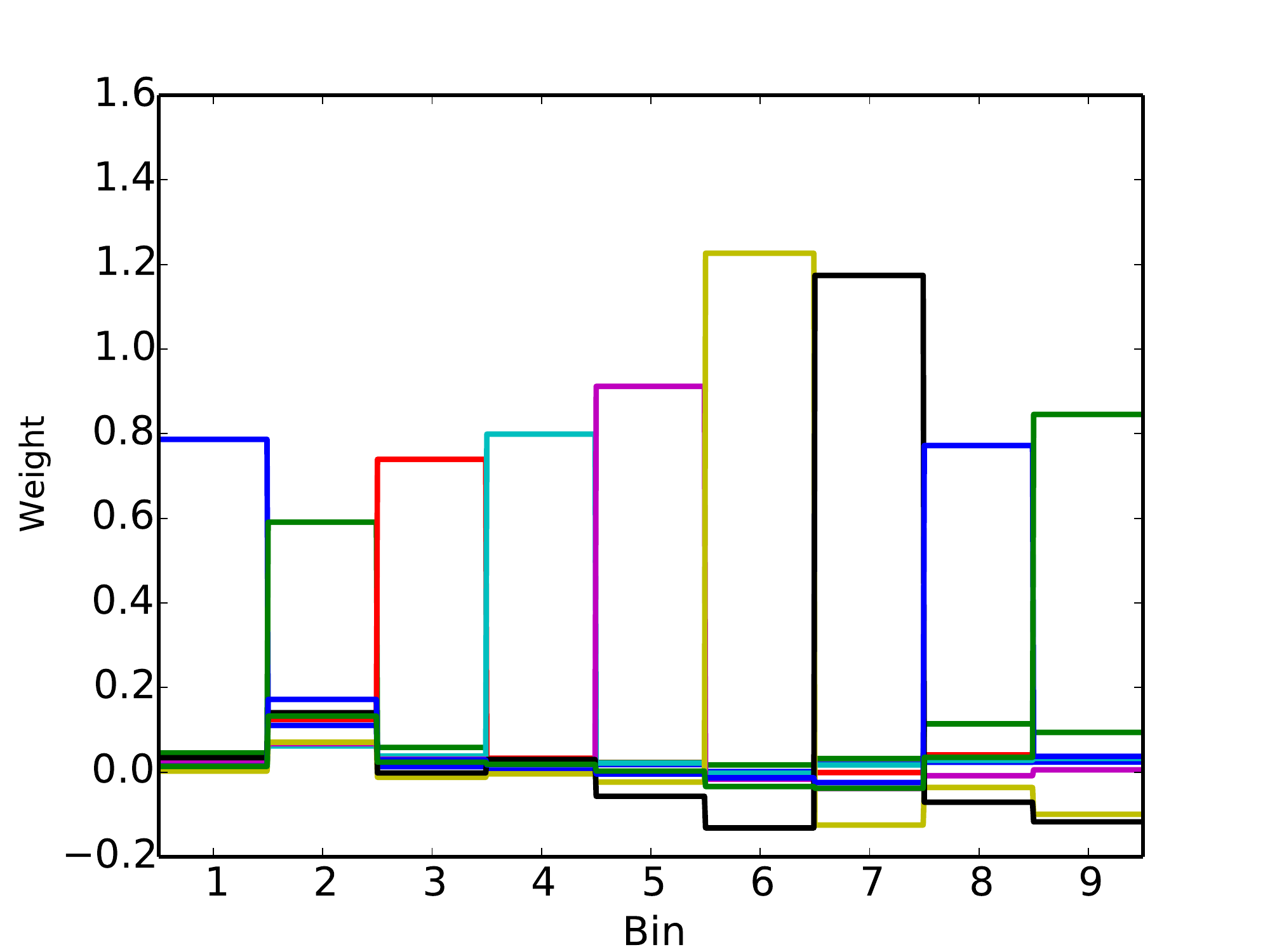}}
\subfigure{\includegraphics[width=3in]{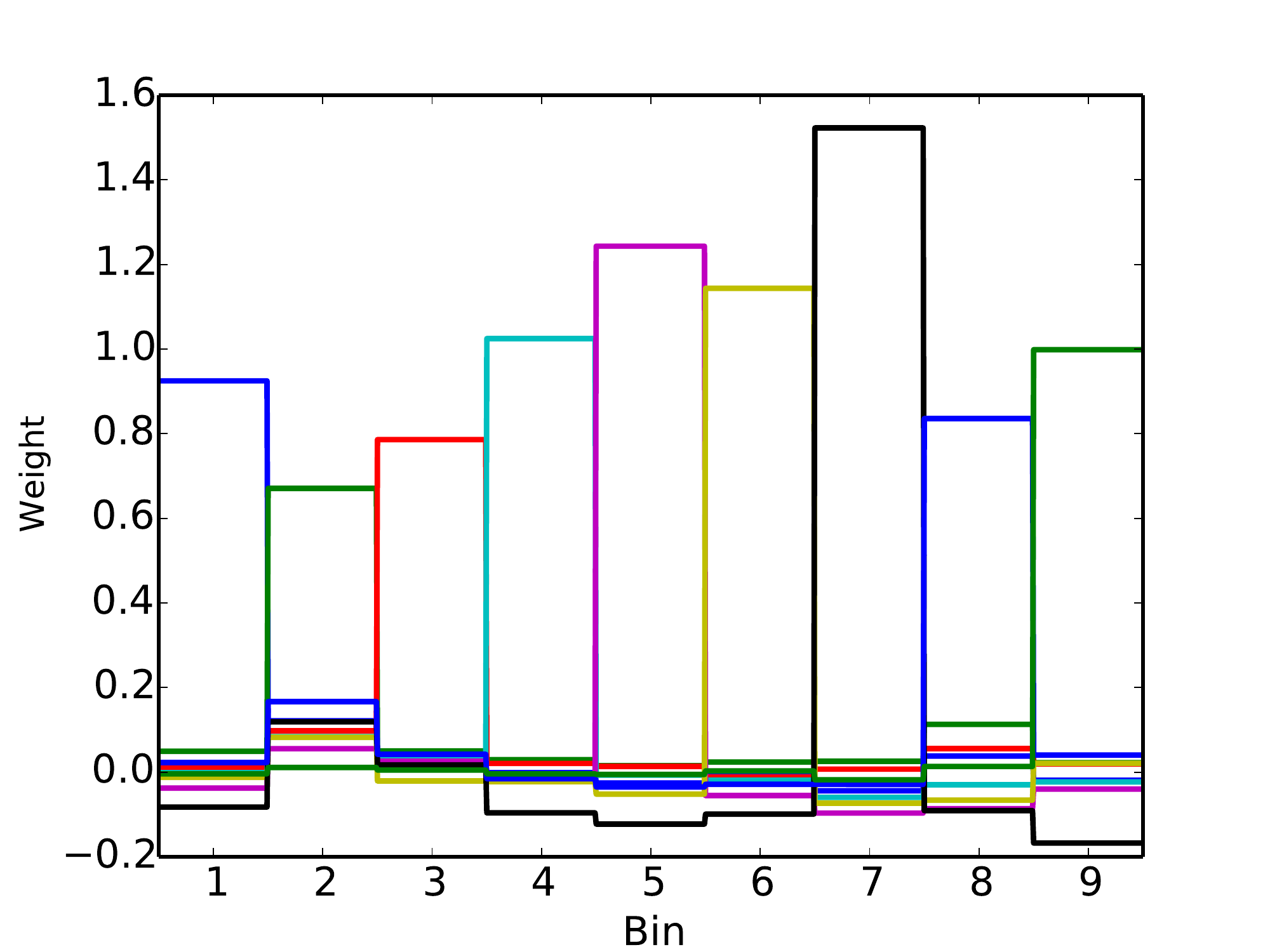}}
\caption{The weight of each mode of LPCA method for WMAP9+SN+BAO(left) and Planck+SN+BAO(right)}\label{figure:weightvalue}
\end{figure}

In Fig. \ref{figure:qvalue}, we show the final $1\sigma $  C.L. constraints on the  uncorrelated parameters(i.e. the parameters $q_i$) representing $\delta x_e(z)$. We also plot the weights that describe going from correlated $p_i$ to the uncorrelated $q_i$ in  Fig. \ref{figure:weightvalue}, here we make the weights for $p_i$ sum to unity, actually the significance is independent of which kind of normalization we have. As shown in Fig. \ref{figure:qvalue}, most bins are consistent with the instantaneous model at $1\sigma $ C.L., while there exists weak hint for the deviation from the instantaneous evolution in redshift around $10.6<z<11$ and $11<z<13$ for WMAP+SN+BAO, when Planck+SN+BAO are used, most bins value are consistent with instantaneous model, except the last bin($13<z<15$), the deviation is about 1$\sigma$ C.L.. However, considering large errors, we need more accurate data to confirm it. In the Fig. \ref{figure:weightvalue}, we plot the weight for each bin, and the weight function well shows its positive and localized properties, which make the uncorrelated parameters approximately one-to-one correspond to original $x_e^i$s and better represents the true $x_e(z)$.

\section{Summary and discussion}\label{Sum}
Reionization optical depth parameter $\tau$ is an important cosmological parameter for CMB, which describes the Thomson scattering between the free streaming CMB photons and free electrons. $\tau$ can be tightly constrained from CMB observations, since the Thomson scattering in the epoch of reionization damps the CMB temperature and polarization spectra at angular scales smaller than the horizon at reionization, and generates E polarization at large angular scale. The reionization history can influence the constraint on $\tau$ for $\tau$ is the integration of the ionized fraction parameter $x_e(z)$, thus the excessive assumption will bias the constraint on $\tau$.

Usually, using the cosmological parameters for constraining $\tau$ will adopt the so called instantaneous reionization model which assume that the reionizing process is very fast. By simulating the future accurate data with a model which is very different from the usually adopted instantaneous model and constraining $\tau$ by adopting an instantaneous model, the final constraints can be highly biased. In order to study the bias on constraining $\tau$ introduced by the instantaneous assumption, we in this paper perform a model independent analysis. We take $x_e(z)$ to be free cosmological parameters and do a global fitting analysis for getting constraints on the evolution of $x_e(z)$ by using the data sets of WMAP, Planck, respectively, combining with BAO and Supernovae. The constraint on each bin is not tight enough, but still we can see that the result is consistent with instantaneous model, however there exists deviation at very few bins. We also adopt the PCA method to get better constraints. We find that, by adopting less noisy modes better constraints can be obtained, and choosing the first 3 modes, it can provide the best reconstructed $x_e(z)$ in which there's weak hints for deviation from an instantaneous evolution.  We hope that, combining with more astrophysical observational data so that we can get more information on the reionization history.

\section*{Acknowledgements}

We acknowledge the use of the Legacy Archive for Microwave
Background Data Analysis (LAMBDA). We thank Antony Lewis, Junqing Xia, Wei Zheng and Gongbo Zhao for helpful
discussion. H. L. is supported in part by the NSFC under Grant No. 11033005 and the youth innovation promotion association project and the Outstanding young scientists project of the Chinese Academy of Sciences. X. Z. are supported in part by the National Science Foundation of China under Grants No. 11121092, No.11375202 and No. 11033005. This paper is supported in part by the CAS pilotB program.


\end{document}